\documentstyle[aps,epsf,multicol]{revtex}

\begin{document}
\draft

\title{Effect of the coupling to a superconductor\\
on the level statistics of a metal grain in a magnetic field}

\author{K. M. Frahm, P. W. Brouwer, J. A. Melsen, and C. W. J. 
Beenakker}

\address{Instituut-Lorentz, University of Leiden\\
P.O. Box 9506, 2300 RA Leiden, The Netherlands}
\date{to appear in Phys. Rev. Lett. {\bf 76}, April (1996)}
\maketitle

\begin{abstract}
A theory is presented for the statistics of the excitation spectrum of 
a disordered metal grain in contact with a superconductor. 
A magnetic field is applied to fully break time-reversal symmetry in 
the grain. Still, an excitation gap of the order of $\delta$ opens up 
provided $N\Gamma^2\gtrsim 1$. Here $\delta$ is the mean level spacing 
in the grain, $\Gamma$ the tunnel probability through the contact
with the superconductor, and $N$ the number of transverse modes in the 
contact region. This provides a microscopic justification for the new 
random-matrix ensemble of Altland and Zirnbauer. 
\end{abstract}

\begin{multicols}{2}
\narrowtext

The proximity to a superconductor is known to induce a gap in the excitation 
spectrum of a normal metal. Semiclassical theories of this ``proximity 
effect'' show that the gap closes if time-reversal symmetry (${\cal T}$) 
is broken (by a magnetic field or by magnetic impurities). Recently, 
Altland and Zirnbauer \cite{zirn} argued that a gap remains in the spectrum 
of a metal grain surrounded by a superconductor --- even if ${\cal T}$ 
is broken completely. (The classical mechanics of such a system had 
previously been studied \cite{kosztin}.) The gap is small 
(of the order of the mean level spacing in the grain), but it has the 
fundamental implication that the level statistics is no longer 
described by the Gaussian unitary ensemble (GUE) of random-matrix theory 
\cite{Mehta}. 

The GUE has a probability distribution of energy levels of the form 
\begin{equation}
\label{eq:gue_dist}
P(\{E_n\})\propto \prod_{i<j} (E_i-E_j)^2
\prod_k \exp\left(-c E_k^2\right),
\end{equation}
with some constant $c>0$ depending on the mean level spacing at the 
Fermi level (chosen at $E=0$). This ensemble was first applied to a granular 
metal by Gorkov and Eliashberg \cite{gorkov}, and derived from microscopic 
theory by Efetov many years later \cite{efetov}. A single-particle 
energy level $E_n$ corresponds to an excitation energy $|E_n|$, that is 
to say, the excitation spectrum is obtained by folding the single-particle 
spectrum along the Fermi level. The folded GUE has been 
studied in Ref. \cite{bruun}. Altland and Zirnbauer introduce a 
different probability distribution,
\begin{equation}
\label{eq:lag_dist}
P(\{E_n\})\propto \prod_{i<j} (E_i^2-E_j^2)^2
\prod_k E_k^2\,\exp\left(-2 c E_k^2\right),
\end{equation}
for the (positive) excitation energies of a metal grain in contact with 
a superconductor. (The excitation spectrum is discrete for $E<\Delta$, with 
$\Delta$ the excitation gap in the bulk of the superconductor.) 
The distribution (\ref{eq:lag_dist}) 
is related to the Laguerre unitary ensemble (LUE) of random-matrix theory 
\cite{slevin} by a change of variables. The density of states $\rho(E)$ in 
this ensemble vanishes quadratically near zero energy \cite{zirn,slevin},
\begin{equation}
\label{eq:lag_dens}
\rho(E)= \frac{1}{\delta}\left(1-\frac{\sin(2\pi E/\delta)}
{2\pi E/\delta}\right).
\end{equation}
The gap in the excitation spectrum is of the order of the mean level 
spacing $\delta$. The folded GUE, on the contrary, has no gap but a 
constant $\rho(E)=1/\delta$ near $E=0$. 

In this paper we present the first microscopic theory for the effect on 
the level statistics of the coupling to a superconductor. 
We consider the case that the conventional proximity effect is fully 
destroyed by a ${\cal T}$-breaking magnetic field \cite{joostetal}. 
Assuming non-interacting quasiparticle excitations, and 
starting from the well-established GUE for the level statistics of an 
isolated metal grain, we obtain a crossover to Altland and Zirnbauer's 
distribution (\ref{eq:lag_dist}) as 
the coupling to a superconductor is increased. 
This provides a microscopic justification for the ``maximum-entropy'' 
hypothesis on which Ref. \cite{zirn} was based. Such a justification is 
needed because, in contrast to ensembles in statistical mechanics, 
there is no physical principle that would require a random-matrix ensemble 
to maximize entropy. Furthermore, because the argument of Ref. 
\cite{zirn} is 
based on the presence or absence of a certain discrete symmetry in the 
Hamiltonian, it can not provide a criterion for how strong the coupling to 
the superconductor should be for the new ensemble to apply. Our microscopic 
approach permits us to identify this criterion, and to compute explicitly 
how the gap in $\rho(E)$ opens up as the coupling strength is increased. 

\begin{figure}
\epsfxsize=0.5\hsize
\epsffile{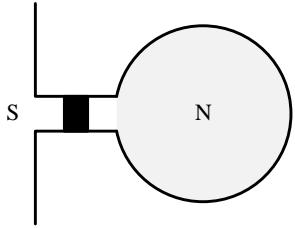}
\caption{\label{fig:dot} A disordered normal-metal grain (N) coupled to 
a superconductor (S). The black area indicates a tunnel barrier.}
\end{figure}

We consider the geometry shown in Fig. \ref{fig:dot} of a disordered 
metal grain (N), which is connected to a superconductor (S) by a point 
contact or microbridge 
containing a tunnel barrier. Breaking ${\cal T}$ requires a 
magnetic field of at most a flux quantum through the grain. This field 
is less than the critical field of the superconductor 
if the size of the grain is greater than the superconducting coherence 
length. For simplicity of presentation we consider a real order parameter 
$\Delta$ in S. (We have found that a spatial dependence of the 
superconducting phase, considered in Ref. \cite{zirn}, has no effect on 
the level statistics in the absence of ${\cal T}$.) We assume zero 
temperature, so that motion in the grain is totally phase coherent. We 
seek the distribution of the excitation energies $E_n\ll \Delta$. We first 
consider the density of states $\rho(E)$.

To determine $\rho(E)$ we adopt the scattering approach of Ref. 
\cite{been1}. We model the point contact by a normal-metal 
lead supporting $N$ transverse modes at the Fermi level. Andreev reflection 
at the interface scatters electrons into holes. This 
corresponds to the off-diagonal blocks in the scattering matrix 
$S_{\rm A}$ for Andreev reflection, 
\begin{equation}
\label{eq:scat_and}
S_{\rm A}=
\left(\begin{array}{cc}
0 & -i \\
-i & 0 \\
\end{array}\right),
\end{equation}
where each of the four blocks is an $N\times N$ matrix. The 
scattering matrix $S_{\rm N}$ for the normal-metal grain plus tunnel 
barrier does not couple electrons and holes, and thus has the block 
diagonal form
\begin{equation}
\label{eq:scat_norm}
S_{\rm N} = \left(\begin{array}{cc} 
S_0(E) & 0 \\ 0 & S_0^*(-E) 
\end{array}\right).
\end{equation}
Here $S_0$ ($S_0^*$) is the scattering matrix for electrons (holes) at 
an energy $E$ from the Fermi level. The $N\times N$ scattering matrix 
$S_0$ can be expressed in terms of the $M\times M$ Hamiltonian $H_0$ of 
the isolated grain and an $M\times N$ coupling matrix $W$ 
\cite{vwz,iwz}: 
\begin{equation}
\label{eq:scat_ham}
S_0(E) = 1-2\pi i W^{\dagger}(E-H_0+i \pi W W^{\dagger})^{-1}W.
\end{equation}
The finite dimension $M$ of $H_0$ is artificial and will be taken to 
infinity later on. 

As demonstrated by Efetov \cite{efetov}, an ensemble of disordered 
metal grains in a magnetic field can be 
described by the GUE \cite{ergodic},
\begin{equation}
\label{eq:gue_dist2}
P(H_0)\propto \exp\left(-c\,\mbox{Tr}\,H_0^2\right).
\end{equation}
(Eq. (\ref{eq:gue_dist}) follows upon integration over the eigenvectors 
of $H_0$.) The coefficient $c$ is related to $\delta$ by 
$c=\pi^2/8M\delta^2$. We recall that $\delta$ is the mean level spacing 
in the folded GUE, which is one half the mean level spacing of $H_0$. 
The coupling matrix $W$ has the form \cite{iwz,pietqdot} 
\begin{eqnarray}
\nonumber
W_{mn} & = & \delta_{mn}\left(\frac{2M\delta}{\pi^2}\right)^{1/2}\!\!
\left(2\Gamma_n^{-1}-1+ 2\Gamma_n^{-1}\sqrt{1-\Gamma_n}\right)^{1/2},\\
\label{eq:coup_def}
&& m=1,2,\ldots,M,\quad n=1,2,\ldots,N. 
\end{eqnarray}
Here $\Gamma_n$ is the tunnel 
probability of mode $n$ through the normal lead \cite{unique}. 
For later use we introduce 
a parameter $\lambda=2M\delta/\pi$ and an $M\times M$ matrix 
$X=(\pi/\lambda) W W^{\rm T}$. In view of Eq.\ (\ref{eq:coup_def}), the matrix 
$X$ is diagonal with $N$ non-zero diagonal elements $x_n$, related to 
$\Gamma_n$ by 
\begin{equation}
\label{eq:Gamma_x_rel}
\Gamma_n=4 x_n\,(1+x_n)^{-2}.
\end{equation}

The excitation energies $E_n$ are the positive roots of the equation 
Det$[1-S_{\rm A}\,S_{\rm N}(E)]=0$,
which can be rewritten as an 
eigenvalue equation \cite{footnote}, 
\begin{equation}
\label{eq:eigen}
\mbox{Det}(E-{\cal H})=0,\quad
{\cal H} = \left(\begin{array}{cc} 
H_0 & -\lambda X \\ 
- \lambda X & -H_0^*
\end{array}\right).
\end{equation}
The effective Hamiltonian ${\cal H}$ is the key theoretical innovation of 
this work. It should not be confused with the Bogoliubov-de Gennes 
Hamiltonian ${\cal H}_{\rm BG}$, which contains the superconducting 
order parameter in the off-diagonal blocks \cite{degennes}. The 
Hamiltonian ${\cal H}_{\rm BG}$ determines the entire excitation spectrum 
(both the discrete part below $\Delta$ and the continuous part above 
$\Delta$), while the effective Hamiltonian ${\cal H}$ determines 
only the low-lying excitations $E_n\ll\Delta$. As we will see, the 
spectrum of ${\cal H}$ can be obtained from a mapping onto a generalization 
of the well-known non-linear $\sigma$-model.
The Hermitian matrix ${\cal H}$ is antisymmetric under the combined 
operation of charge conjugation (${\cal C}$) and time inversion 
(${\cal T}$): 
\begin{equation}
\label{eq:ct_sym}
{\cal H}=-{\cal C}^{\rm T} {\cal H}^{\rm T}{\cal C},\quad
{\cal C}=
\left(\begin{array}{cc}
0 & -1 \\
1 &  0 \\
\end{array}\right). 
\end{equation}
The ${\cal CT}$-antisymmetry ensures that the eigenvalues of 
${\cal H}$ lie symmetrically around $E=0$. This 
discrete symmetry (for ${\cal H}_{\rm BG}$) was the main 
point in the maximum-entropy argument of 
Altland and Zirnbauer \cite{zirn}. 

To compute the spectral statistics on the scale of the level spacing, 
we need a non-perturbative technique. We employ the supersymmetric method 
\cite{efetov,vwz}, suitably modified \cite{andreev2} to incorporate the 
special symmetry (\ref{eq:ct_sym}) of ${\cal H}$. The density of states 
\begin{equation}
\label{}
\rho(E)=\frac{1}{2\pi}\lim_{z\to 0} \frac{d}{d z} 
\,\mbox{Im}\,F(z)
\end{equation}
is obtained from the generating function
\begin{mathletters}
\label{eq:gen_func}
\begin{eqnarray}
\label{eq:gen_func_main}
&& F(z)=\int d\tilde\phi\left\langle\exp[
{\textstyle \frac{1}{2}}i\tilde \phi^\dagger(E+i\epsilon -
\tilde{\cal H}+z L)\tilde \phi]\right\rangle,\\
\label{eq:mat1}
&&\tilde\phi=
\left(\begin{array}{c}
\phi  \\
{\cal C}\phi^*  \\
\end{array}\right),\quad
\tilde{\cal H}=
\left(\begin{array}{cc}
{\cal H}\, \openone_2 & 0 \\
0 & -{\cal H}\, \openone_2  \\
\end{array}\right),\quad\\
\label{eq:mat2}
&&L=\left(\begin{array}{cc}
\openone_{2M}\,\sigma_3 & 0 \\
0 & \openone_{2M}\,\sigma_3 \\
\end{array}\right).
\end{eqnarray}
\end{mathletters}%
Here $\phi$ is a $4M$-component supervector 
containing $2M$ commuting and $2M$ anticommuting variables. 
Half of each $2M$ variables correspond to electron states and half to 
hole states. The charge conjugation operator 
${\cal C}$ interchanges electron and hole variables. The matrices 
${\cal H}\,\openone_2$ and $\openone_{2M}\,\sigma_3$ 
are tensor products between a $2M\times 2M$ and a $2\times 2$ matrix 
($\openone_p$ is the $p$-dimensional unit matrix and $\sigma_3$ is a 
Pauli matrix). The appearance of 
$-{\cal H}$ in the ${\cal CT}$-conjugated block of $\tilde{\cal H}$ reflects 
the ${\cal CT}$-antisymmetry (\ref{eq:ct_sym}) of ${\cal H}$. The measure 
$d\tilde\phi$ is normalized such that $F(0)=1$. The brackets 
$\langle\cdots\rangle$ indicate an average over $H_0$ with 
distribution (\ref{eq:gue_dist2}). 

To evaluate $F(z)$ we perform a series of steps which are by now standard 
in the field \cite{efetov,vwz,andreev2}. 
We first average $H_0$ over the GUE, which can be done exactly since it 
involves only Gaussian integrals. A term which is quartic in $\tilde\phi$ 
appears, and we decouple it by a Hubbard-Stratonovich transformation. 
This transformation introduces an additional integral over an $8\times 8$ 
supermatrix $Q$, which we evaluate by a saddle-point approximation that 
becomes exact in the limit $M\to\infty$. We solve the saddle-point equation 
in the limit $E\to 0$ at fixed $N$ and $E/\delta$. 
As in Ref. \cite{andreev2}, a manifold of saddle points 
(determined by $Q^2=1$) appears in this limit, 
while for $E\gg\delta$ only a single saddle point remains. 

The matrices $Q$ on the saddle-point manifold have the electron-hole 
block structure 
\begin{equation}
\label{eq:q_block_struc}
Q=
\left(\begin{array}{cc}
Q_1 & 0 \\
0  & Q_2 \\
\end{array}\right),\quad
Q_2=-\tilde{\cal C}^{\rm T} Q_1^{\rm T}\tilde{\cal C},\quad
\tilde{\cal C}=
\left(\begin{array}{cc}
0 & \sigma_3 \\
\openone_2 & 0 \\
\end{array}\right). 
\end{equation}
The $4\times 4$ supermatrix $Q_1$ belongs to the 
coset space of the non-linear $\sigma$-model in the unitary symmetry 
class, and $-Q_2$ is the ${\cal CT}$-conjugate of $Q_1$. The 
matrix $\tilde{\cal C}$ is the charge conjugation operator for the 
$\sigma$-model. The density of states is obtained as an 
integral over the saddle-point manifold, 
\begin{mathletters}
\label{eq:rho_qint_tot}
\begin{eqnarray}
\nonumber 
\rho(E) & = & \mbox{Im}\biggl\{\frac{i}{8\delta}\int dQ_1
\ \mbox{Str}[\tilde L\tau(Q_1+Q_2)]\\
\label{eq:rho_qint}
&&\times \exp[-{\cal L}_1(Q_1)-{\cal L}_2(Q_1)]\biggr\},\\
\label{eq:l1_def}
{\cal L}_1(Q_1) & = & -\frac{i\pi}{4\delta}\,(E+i\epsilon)
\ \mbox{Str}[\tau(Q_1+Q_2)],\\
\label{eq:l2_def}
{\cal L}_2(Q_1) & = & \frac{1}{2}\sum_{j=1}^N \mbox{Str}
\ [\ln(1+x_j^2 Q_1 Q_2)],
\end{eqnarray}
\end{mathletters}%
where Str denotes the supertrace and 
\begin{equation}
\tilde L  =  \left(\begin{array}{cc}
\sigma_3 & 0  \\
0 & \sigma_3 \\
\end{array}\right),\quad
\tau=\left(\begin{array}{cc}
\openone_2 & 0  \\
0 & -\openone_2 \\
\end{array}\right). 
\end{equation}

The action ${\cal L}_2$ can be simplified by expanding it in powers 
of $Q_1-Q_2$. This is justified either if $\Gamma_n\ll 1$ for all $n$ 
or if $N\gg 1$. (We therefore exclude the case that $N$ and $\Gamma_n$ 
are both close to $1$.) 
The first non-vanishing term in this expansion is 
\begin{eqnarray}
\label{eq:l2_sim}
{\cal L}_2(Q_1) & = & -\case{1}{32}\,g_{\rm A}\ \mbox{Str}[(Q_1-Q_2)^2],\\
\label{eq:ga_def}
g_{\rm A} & = & \sum_{j=1}^N\frac{8x_j^2}{(1+x_j^2)^2}=
\sum_{j=1}^N \frac{2\Gamma_j^2}{(2-\Gamma_j)^2}.
\end{eqnarray}
The parameter $g_{\rm A}$ is the Andreev conductance \cite{blonder,been2}
of the tunnel barrier at the NS interface, which can be much smaller than the 
normal-state conductance $g=\sum_{j=1}^N \Gamma_j$. 
(Both conductances are in units of $2e^2/h$.) For identical tunnel 
probabilities $\Gamma_j\equiv\Gamma\ll 1$ one has $g=N\Gamma$ while 
$g_{\rm A}=\frac{1}{2}\,N\Gamma^2$. 

\begin{figure}
\epsfxsize=0.8
\hsize
\epsffile{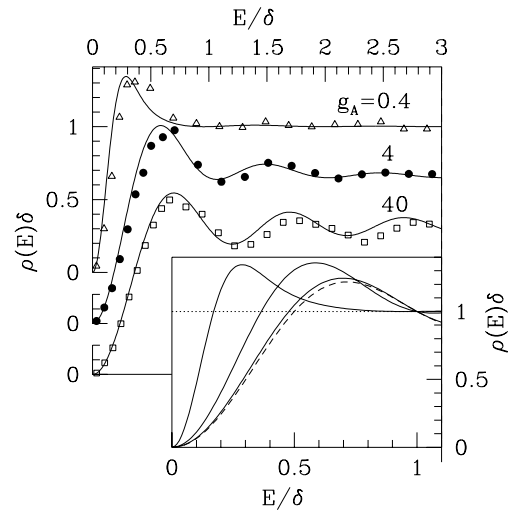}
\caption{\label{fig:dens} 
Density of states for three different values of $g_{\rm A}=
0.4,\ 4,\ 40$. The solid curves are the analytical result 
(\protect\ref{eq:rho_result1}), the data points are from a numerical 
solution of Eq. (\protect\ref{eq:eigen}) [with $N=20$, $M=100$, and 
a mode-independent tunnel probability $\Gamma_j\equiv\Gamma$ 
determined by Eq. (\protect\ref{eq:ga_def}); some $10^4$ 
random matrices $H_0$ in the GUE where generated to compute $\rho(E)$].
In the inset the analytical result is shown on an expanded scale for the 
same values of $g_{\rm A}$ as in the main plot. The dashed line 
is Eq. (\protect\ref{eq:lag_dens}), corresponding to the 
limit $g_{\rm A}\to\infty$. The dotted line corresponds to the 
limit $g_{\rm A}\to 0$ of a folded GUE. }
\end{figure}

Finally, we evaluate the integral (\ref{eq:rho_qint_tot}) using the standard 
decomposition of $Q_1$ in terms of angular 
and radial variables \cite{efetov,vwz,andreev2}. The result is
\begin{equation}
\label{eq:rho_result1}
\rho(E)=\frac{1}{\delta}
-\frac{\sin(\pi E/\delta)}{\pi E}\int_0^\infty ds
\ e^{-s}\,\cos\left(\frac{\pi E}{\delta}\sqrt{1+\frac{4s}{g_{\rm A}}}
\,\right).
\end{equation}
Eq.\ (\ref{eq:rho_result1}) describes the crossover from $\rho(E)=1/\delta$ 
for $g_{\rm A}\ll 1$ to Altland and Zirnbauer's result (\ref{eq:lag_dens}) 
for $g_{\rm A}\gg 1$. In Fig. \ref{fig:dens} we have plotted the opening 
of the gap as the coupling to the superconductor is increased. The 
${\cal CT}$-symmetry becomes effective at an energy $E$ for 
$g_{\rm A}\gtrsim E/\delta$. 
For small energies $E\ll \delta\,\min (\sqrt{g_{\rm A}},\,1)$ 
the density of states 
vanishes quadratically, regardless of how weak the coupling is. 

As a check on our calculations, we have also computed $\rho(E)$ 
numerically from the eigenvalue equation (\ref{eq:eigen}), by generating 
a large number of random matrices $H_0$ in the GUE. The numerical results 
(data points in Fig. \ref{fig:dens}) are in good agreement with Eq. 
(\ref{eq:rho_result1}). 

The parameter $g_{\rm A}$ which governs the opening of the excitation gap, 
does so by enforcing a ${\cal CT}$-antisymmetry on the non-linear 
$\sigma$-model. To see this, consider the term (\ref{eq:l2_sim}) in the 
action, which 
is proportional to $g_{\rm A}$. For $g_{\rm A}\gg 1$ this term constrains 
$Q_2$ to be close to $Q_1$, and in the limit $g_{\rm A}\to\infty$ one 
obtains the ${\cal CT}$-antisymmetry 
\begin{equation}
\label{eq:ct_q_sym}
Q_2=-\tilde{\cal C}^{\rm T}\,Q_1^{\rm T}\,\tilde{\cal C}=Q_1.
\end{equation}
For $g_{\rm A}\ll 1$, on the contrary, $Q_2$ may be quite different from 
$Q_1$, and the ${\cal CT}$-antisymmetry is effectively broken. 

We generalized these considerations to level-density correlation functions. 
For this, one has to consider a more 
general source term [replacing the term $z\,L$ in Eq. (\ref{eq:gen_func})],
and higher-dimensional 
supervectors (containing both advanced and retarded components). 
After carrying out the same steps outlined above for the density of 
states, we arrive at a non-linear $\sigma$-model with a broken 
${\cal CT}$-antisymmetry. This symmetry is restored for $g_{\rm A}\to\infty$, 
when the $\sigma$-model becomes equivalent to that associated with the 
Laguerre unitary ensemble of Ref. \cite{zirn}. This 
establishes the validity of the distribution (\ref{eq:lag_dist}) in the 
limit of a strong coupling to the superconductor. 

In summary, we have presented a microscopic theory for the random-matrix 
ensemble which Altland and Zirnbauer obtained from a maximum-entropy 
hypothesis. The ${\cal CT}$-antisymmetry of the Hamiltonian of 
non-interacting quasiparticles induces an excitation gap even if the 
conventional proximity effect is destroyed by a magnetic field. The 
Andreev conductance $g_{\rm A}\simeq \frac{1}{2}\,N\Gamma^2$ 
of the contact between 
the normal metal and the superconductor governs the size of the gap, 
which becomes of the order of the mean level spacing $\delta$ for 
$g_{\rm A}\gg 1$. An interesting problem for future research 
\cite{altshuler} is the sensitivity of the gap to Coulomb interactions 
between the quasiparticles, which break the charge-conjugation invariance 
of the Hamiltonian. 

This work was supported by the Dutch Science Foundation NWO/FOM 
and by the Human Capital and Mobility program of the European Community. 


\end{multicols}


\begin{references}
\bibitem{ergodic} The GUE holds on energy scales smaller than the 
inverse ergodic time $\hbar/t_{\rm erg}$ (where $t_{\rm erg}=L^2/D$ 
is the time in which an electron with diffusion constant $D$ diffuses 
across a grain of size $L$). On this energy scale the density of states 
of the isolated grain is constant and indistinguishable from the semi-circular 
density of states of the GUE. In a metal $\hbar/t_{\rm erg}$ is much greater 
than the mean level spacing $\delta$, so that we can stay far from 
the non-universal regime.
\bibitem{unique} The set of tunnel probabilities $\{\Gamma_n\}$ does not 
determine the coupling matrix $W$ uniquely.\ Firstly, we have the freedom 
to perform an orthogonal transformation $W\to O_1\,W\,O_2$, with 
$O_1$ and $O_2$ orthogonal matrices. (We assume that ${\cal T}$ is not 
broken on the length scale of the tunnel barrier, which is why 
we take orthogonal --- rather than unitary --- transformations.) 
Secondly, we have the freedom to chose the sign of the square root 
in Eq. (\ref{eq:coup_def}): $\sqrt{1-\Gamma_n}\to -\sqrt{1-\Gamma_n}$. 
The distribution of the excitation energies is invariant under 
both types of transformations.
\bibitem{footnote} According to Ref. \cite{been1} the discrete spectrum 
is determined by Det$[1-\alpha(E)\,S_{\rm A}\,S_{\rm N}(E)]=0$, with 
$\alpha(E)=i\,\exp[-i\,\mbox{arccos}(E/\Delta)]$. We may replace $\alpha(E)$ 
by $1$ if both $E$ and $\hbar/t_{\rm dwell}$ are $\ll\Delta$, where 
$t_{\rm dwell}=\hbar(\delta\,\sum_j \Gamma_j)^{-1}$ is the mean dwell 
time of an electron in the grain. 
More generally, if $E\ll\Delta$ but $\hbar/t_{\rm dwell}$ arbitrary, our 
main result (\ref{eq:rho_result1}) remains valid if we 
replace $\delta\,\to\,\delta_{\rm eff}$ with 
$\delta_{\rm eff}^{-1}=\delta^{-1}+(\pi\Delta)^{-1}\,
\sum_j\Gamma_j\,(2-\Gamma_j)^{-1}$.
\end{references}
\end{document}